\documentclass[aps,prb,twocolumn,groupedaddress,showpacs]{revtex4}
\usepackage[dvips]{graphicx}
\newcommand{\Vec}[1]{\mbox{\boldmath$#1$}}

\begin{document}


\title{Large orbital magnetic moments in carbon nanotubes generated by 
resonant transport}

\author{Naoto Tsuji, Shigehiro Takajo$^*$ and Hideo Aoki}
\affiliation{Department of Physics, University of Tokyo, Hongo 7-3-1, 
Bunkyo-ku, Tokyo 113-0033, Japan}


\date{\today}

\begin{abstract}
The nonequilibrium Green's function method is used to study the 
ballistic transport in metallic carbon nanotubes when a current 
is injected from the electrodes with finite 
bias voltages.  We reveal, both analytically and 
numerically, that large loop currents circulating around the 
tube are induced, which come from a quantum-mechanical 
interference and are much larger than the current along the tube axis 
when the injected electron is resonant with a time-reversed 
pair of degenerate states, which are, in fact, inherent in the 
zigzag and chiral nanotubes.  
The loop current produces large orbital magnetic moments, 
making the nanotube a molecular solenoid.  

\end{abstract}



\maketitle

\section*{INTRODUCTION} 
Carbon nanotubes (CNTs) have remarkable electronic properties 
related with their unique geometrical structure.\cite{Saito,Reich} 
For example, the way in which the graphene sheet is wound 
into a cylinder determines whether CNTs are metals or 
semiconductors.\cite{Hamada,Saito2} 
Recent advances have made the measurements of electrical properties of 
individual single-wall CNTs possible.\cite{Bockrath,Tans} 
Most recently, orbital magnetic moments of a single-wall CNT have been 
detected, where the presence of moment is deduced 
from the shift of energy levels in external magnetic fields.\cite{Minot,Coskun,Zaric} 
Since the magnitude of magnetic moments estimated in the experiment is 
about ten times the Bohr magneton, the CNT as a molecular solenoid 
has attracted attention. 

Semiclassically, the large orbital magnetic moments can be understood as an 
effect of chiral currents, for which several semiclassical calculations 
have been done based 
on Boltzmann's equation.\cite{Miyamoto,Yevtushenko} 
However, the approach, being applicable when the 
transport is diffusive and lacks coherence, 
cannot treat purely quantum mechanical effects.  
Quantum mechanical 
treatments of magnetic properties of CNTs have previously been 
given,\cite{Ajiki-Ando,Ando,Lu,Marganska} 
but those studies have concentrated on the energy shift and 
moment of an 
isolated CNT in external magnetic fields in equilibrium, 
so the currents and the effects of electrodes have not been studied. 
Since ballistic transport can be strongly affected by 
electrodes, we need to take account of the influence of electrodes 
in CNTs, which is exactly our motivation here. 
In the process, the quantum loop 
current, a purely quantum mechanical effect, has turned out to 
be indeed large, which we study by making 
use of the nonequilibrium Green's function method.\cite{Datta} 

A loop current has been studied in the context of 
small molecules placed between the scanning tunneling microscope 
(STM) tip and the substrate.  
Specifically, Nakanishi and Tsukada\cite{Nakanishi} 
and Tsukada {\it et al.}\cite{Tsukada}
have shown that a loop current
can be dramatically amplified when the energy of the injected 
electron is resonant with degenerate eigenstates of
an isolated molecule.  This mechanism should also be applicable to CNTs, 
for which a special interest 
is the effect of inherent degeneracies associated with wave functions 
traveling clockwise and anticlockwise around the tube. 
We develop a perturbation expansion of the nonequilibrium Green's function 
in the weak coupling between the conductor and electrodes.  
This serves to identify how the loop current paths are determined 
in terms of phase variation of wave functions.  
The peculiarity of the CNTs appears as their band structure that can satisfy 
the condition for large orbital magnetic moments.

We then numerically calculate the current distribution and the magnitude 
of the orbital magnetic 
moments as a function of the bias voltage, which confirms the analytic discussions. 
The order of magnitude of the 
magnetic moment obtained in the numerical calculation 
is consistent with the value estimated in the experiment.\cite{Minot} 
Throughout this Brief Report, we consider 
metallic CNTs in the absence of external magnetic fields.  
CNTs are characterized by the chiral index
$(n_1,n_2)$ in standard literatures,\cite{Saito,Reich} where 
CNTs are metals when $n_1-n_2\equiv 0\;({\rm mod}3)$ or 
semiconductors otherwise.\cite{Hamada,Saito2} 
Here we concentrate on the one-body problem.  

\section*{FORMALISM} 
\begin{figure}
\includegraphics[width=8cm]{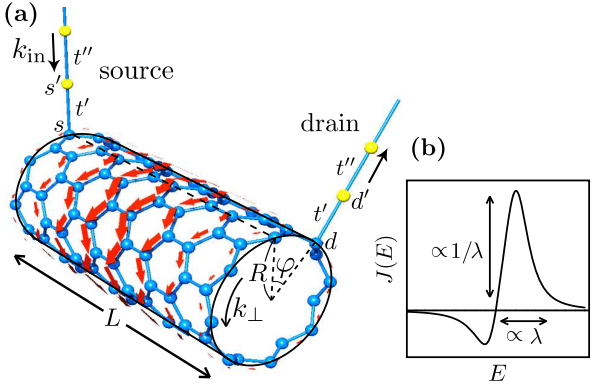}
\caption{\label{cylinder}(Color online) (a) A nanotube attached to 
1D electrodes is schematically shown. The arrows represent the quantum 
loop current. (b) A typical asymmetric peak of the current 
density $J_{ij}^\alpha$ vs energy.
}
\end{figure}

We employ the tight-binding model whose Hamiltonian is 
\begin{eqnarray}
H&=&H_{\rm CNT}+H_{\rm CNT-electrode}+H_{\rm electrode},\nonumber \\
H_{\rm CNT}&=&-t\sum_{\langle ij\rangle\in {\rm CNT}}(c_i^\dagger 
 c_j+c_j^\dagger c_i),\label{hamiltonian1}\\
H_{\rm CNT-electrode}&=&-t'\sum_{p=s, d}(c_p^\dagger c_{p'}+c_{p'}^\dagger c_p),\\
H_{\rm electrode}&=&-t''\sum_{\langle ij\rangle\in {\rm electrode}}(c_i^\dagger 
 c_j+c_j^\dagger c_i).
\label{hamiltonian3}
\end{eqnarray}
Here, $H_{\rm CNT}$ is the tight-binding Hamiltonian for the single-wall 
CNT, 
$H_{\rm CNT-electrode}$ describes the connection between the CNT and the electrodes, 
$H_{\rm electrode}$ describes the electrodes, 
$s,s',d,d'$ are the sites connecting the CNT and the electrodes and 
$t,t',t''(>0)$ are the respective transfer integrals [see Fig. \ref{cylinder}(a)]. 
$t$, the transfer integral of CNT, is $O(1$ eV)\cite{Saito}, while 
we have taken $t''$ for the electrodes to be large enough so that 
their density of states is nearly constant (wide-band limit) 
and $t'$, the hopping 
between CNT and electrodes, to be $\ll t, t''$.
Here, we follow other theoretical literature in assuming electrodes to 
be one dimensional for simplicity. 

For one-dimensional (1D) electrodes, the retarded and advanced Green's functions 
$G^{R,A}(E)=(E-H\pm i\epsilon)^{-1}$ 
[where $+\;(-)$ is assigned to $R\;(A)$] can be cast into a 
form\cite{Datta} 
\begin{eqnarray*}
G^{R,A}(E)&=&(E-H_{\rm CNT}-\Sigma^{R,A})^{-1},\\
\Sigma^{R,A}&=&\Sigma_{s}^{R,A}+\Sigma_{d}^{R,A},\\
\Sigma_p^{R,A}&=&-\lambda t\;e^{\pm ik_{\rm in} a_0}c_p^\dagger c_p\;\;
(p={s,d}),
\end{eqnarray*}
where $\lambda=(t')^2/(tt'')$ characterizes the sample-electrode 
coupling, and $k_{\rm in}$ the 
Bloch's wave number for the incident electron 
having an energy $E=-2t''\;\cos(k_{\rm in} a_0)$ with $a_0$ the lattice 
constant of the conductor and electrodes.  
This reduces the problem to the sample alone, with 
the effects of electrodes contained in the self-energy $\Sigma^{R,A}$.  


We can perform perturbation expansions 
in terms of the weak coupling $\lambda$. 
Since nonperturbative effects often 
arise in resonance phenomena, 
here we adopt the eigenfunction expansion 
method\cite{Datta} that can take nonperturbative corrections 
into account.  
In this formalism, 
we express the Green's function as
$G^R(\Vec{r}_1,\Vec{r}_2)=\sum_\alpha 
\phi_\alpha^R(\Vec{r}_1){\phi_\alpha^A}^\ast(\Vec{r}_2)/(E-E_\alpha)$, 
where $\phi_\alpha^{R,A}(\Vec{r})$ are 
the eigenfunctions  
of the non-Hermitian operators $H_{\rm CNT}+\Sigma^{R,A}$ 
that form a biorthonormal set.
The current $I_{ij}$
flowing from site $i$ to $j$ is given by $I_{ij}=\int_{\mu_d}^{\mu_s}dE
J_{ij}(E)$, where 
$\mu_s$ and $\mu_d$ are the chemical potentials of the source and 
drain electrodes, respectively.
The current density $J_{ij}$ per unit energy is calculated as 
\begin{eqnarray}
\label{J1}
J_{ij}(E)=\frac{4e}{h}{\rm Im}[H_{ij}G^{\rm n}_{ji}(E)],
\end{eqnarray}
where $G^{\rm n}$ is the nonequilibrium Green's function defined by 
the usual Green's functions as 
$G^{\rm n}=G^R\Sigma^{\rm in}G^A$ with 
$\Sigma^{\rm in}=i(\Sigma_s^R-\Sigma_s^A)$ at zero temperature.\cite{Datta,Nakanishi} 
These currents generate an orbital magnetic moment 
$\Vec{M}=\frac{1}{2}\int 
d\Vec{r}\;\Vec{r}\times\Vec{j}(\Vec{r})=\frac{1}{2}\sum_{\langle ij\rangle}I_{ij}\Vec{r}_i\times\Vec{r}_j$.

Let us consider the case where the incident electron is resonant
to doubly degenerate states $\phi_1,\phi_2$ with an eigenenergy 
$\varepsilon_\alpha$.  
In the degenerate perturbation theory, 
the zeroth-order $\phi_1^R\;(=\phi_1^A)$ and $\phi_2^R\;(=\phi_2^A)$, 
having $E_{1,2}=\varepsilon_\alpha+\Sigma_{1,2}^R$ 
with $\Sigma_{1,2}^R$ the eigenvalues of $2\times 2$ $\Sigma^R$, 
can be written as $(\phi_1\pm\phi_2)/\sqrt{2}$, respectively, where 
we fix $\phi_{1,2}$ as the 
{\it time-reversed pair} with $\phi_1=\phi_2^\ast\equiv\phi_\alpha$
and choose the overall phase of $\phi_\alpha$ 
so that $[\phi_\alpha(s)]^2+[\phi_\alpha(d)]^2$ be real. 
The Green's function becomes 
\begin{eqnarray}
&&G^R(i,j)\simeq \frac{\phi_1^R(i){\phi_1^A}^\ast(j)}{E-E_1}+\frac{\phi_2^R(i){\phi_2^A}^\ast(j)}{E-E_2}\hspace{2cm}\nonumber\\
\label{G2}
&=&2\left(\frac{{\rm Re}[\phi_\alpha(i)]{\rm Re}[\phi_\alpha(j)]}{E-\varepsilon_\alpha-\Sigma_1^R}+
 \frac{{\rm Im}[\phi_\alpha(i)]{\rm Im}[\phi_\alpha(j)]}{E-\varepsilon_\alpha-\Sigma_2^R}\right).
\end{eqnarray}
Here, we have only retained the term related to the resonant 
states $\alpha$, which is valid as long as 
the splitting, $\sim \lambda t$, 
of the degenerate levels due to the self-energy 
is smaller than the interval across the adjacent energy level 
[$\sim$ 1 meV$/L(\rm \mu m)$ for CNTs\cite{Reich}], i.e., $\lambda\ll 10^{-3}/L$($\rm \mu m$) 
for CNTs with $t\sim 1\;$eV.

From Eqs. (\ref{J1}) and (\ref{G2}), we end up with
\begin{eqnarray*}
J_{ij}^\alpha(E)&\simeq&\frac{16e}{h}{\rm Im}\{[\phi_\alpha(s)]^2\}\nonumber\\
&\times&\hspace{-.2cm}\frac{\lambda 
 t^2(E-\varepsilon_\alpha){\rm Im}(\Sigma_1^R-\Sigma_2^R)}{|E-\varepsilon_\alpha-\Sigma_1^R|^2|E-\varepsilon_\alpha-\Sigma_2^R|^2}
{\rm Im}[\phi_\alpha(i)\phi_\alpha^\ast(j)].
\end{eqnarray*}
This expression can also be derived by using the results 
of Nakanishi and Tsukada.\cite{Nakanishi} 
We can see that 
the $E$ dependence of the current density $J_{ij}^\alpha$ 
is mainly determined by the factor $(E-\varepsilon_\alpha)/
(|E-\varepsilon_\alpha-\Sigma_1^R|^2|E-\varepsilon_\alpha-\Sigma_2^R|^2)$.
This function exhibits an asymmetric peak 
[with width $\sim \lambda$ and height $\sim 1/\lambda$, see Fig. \ref{cylinder}(b)], 
which is why the integration across the peak results in a finite net current. 
Sasada and Hatano\cite{Hatano} has interpreted the 
asymmetric resonance 
shape for the transmission coefficient as Fano effect, 
i.e., an interference between continuous states in 
electrodes and discrete states in a conductor.  
So the present case may be called a {\it Fano effect extended to loop currents}.  

\section*{ANALYTIC EXPRESSION FOR THE CURRENT AND MAGNETIC MOMENT} 
When we integrate $J_{ij}^\alpha(E)$ to obtain $I_{ij}^\alpha$, we can replace the 
range $\mu_d\leq E\leq\mu_s$ with $-\infty\leq E\leq 
\infty$, since $J_{ij}^\alpha$ is negligible outside the peak.  
By taking care of $E$ dependences in $k_{\rm in}$ and $\Sigma_{1,2}^R$ 
as well, we obtain
\begin{eqnarray}
I_{ij}^\alpha&=&\frac{8e}{\hbar}\frac{t^2\varepsilon_\alpha}{t''}{\rm Im}\{[\phi_\alpha(s)]^2\}
 \frac{\lambda\;{\rm Im}[\Sigma_1^R(\varepsilon_\alpha)-\Sigma_2^R(\varepsilon_\alpha)]}
 {|\Sigma_1^R(\varepsilon_\alpha)-{\Sigma_2^R}^\ast(\varepsilon_\alpha)|^2}\nonumber\\
&&\hspace{1.5cm}\times{\rm Im}[\phi_\alpha(i)\phi_\alpha^\ast(j)]+O(\lambda).
\label{I1}
\end{eqnarray}

\begin{figure}
\includegraphics[width=8cm]{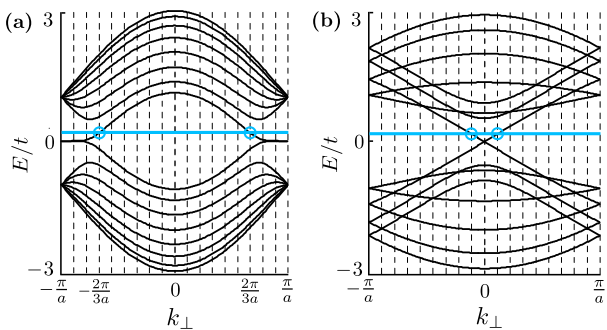}
\caption{\label{band}
(Color online) Band structure against $k_\perp$ for typical (a) zigzag or 
(b) armchair CNTs. 
The vertical dashed lines indicate discrete $k_\perp$ points.  
The horizontal line represents the energy of the incident electron.}
\end{figure}

Since $\Sigma^R\sim \lambda$, 
the leading term of $I_{ij}^\alpha$ is $\sim \lambda^0$. 
This is to be contrasted with a nondegenerate resonant state $\alpha$, 
for which the corresponding expression involves the factor ${\rm 
Im}[\phi_\alpha(i)\phi_\alpha^\ast(j)]$, 
but this must vanish in the absence of external magnetic fields, so 
the current is small [$O(\lambda)$]. 
For the degenerate case, $I_{ij}^\alpha \propto {\rm 
Im}[\phi_\alpha(i)\phi_\alpha^\ast(j)]$ 
indicates that currents flow 
along the direction in which the phase of $\phi_\alpha$ 
varies.   So this implies that a cylindrical conductor, with a 
standing wave along the axis and a propagating wave around the tube, 
should have currents circulating around the tube that is {\it much larger} than 
the component flowing along the axis. 

From Eq. (\ref{I1}), 
the magnetic moment along the tube axis generated by the current becomes
\begin{eqnarray}
(\Vec{M}_{\alpha})_z &=& \frac{2e}{\hbar}
\frac{t}{t''}\varepsilon_\alpha
\frac{{\rm Im}\{[\phi_\alpha^\ast(s)\phi_\alpha(d)]^2\}}{[|\phi_\alpha(s)|^2+|\phi_\alpha(d)|^2]^2}
 \nonumber\\
\label{M2}
&\times&\sum_{\langle ij\rangle}{\rm Im}[\phi_\alpha(i)\phi_\alpha^\ast(j)](\Vec{r}_i\times\Vec{r}_j)_z
  + O(\lambda).
\end{eqnarray}
The total magnetic moment is the sum over the resonant 
states which are relevant to transport. 
While 
$M_\alpha$ is proportional to the resonant energy level 
$\varepsilon_\alpha$, this is only valid for 1D electrodes 
with Eq. (\ref{hamiltonian3}), so the specific form should depend on the detail of the electrodes.  
In addition, we note that, while one might expect that 
edge states, which are known to exist in a finite carbon nanotube 
with a flat dispersion on the Fermi energy,\cite{Fujita} 
may have an important effect on the magnetic moment, 
the contribution from edge states is negligible for 1D electrodes 
since $M_\alpha \propto \varepsilon_\alpha$, as shown in Eq. (\ref{M2}).  
\begin{figure}[htbp]
\begin{center}
\includegraphics[width=7cm]{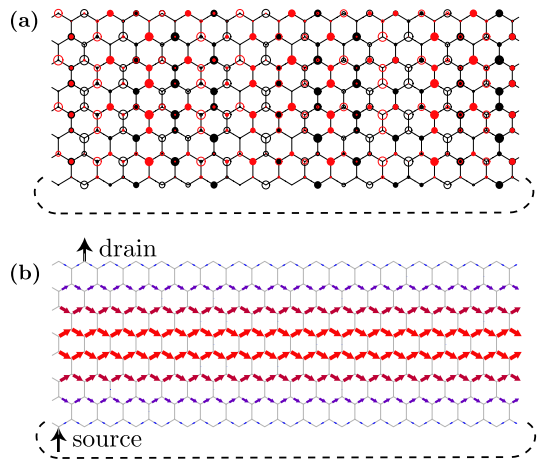}
\caption{\label{zigzag}(Color online) (a) Typical doubly degenerate (in black and red)
eigenfunctions with an eigenenergy of $0.419\;522t$ 
in a zigzag CNT with $(n_1,n_2)=(18,0)$ and $L=11a_0/\sqrt{3}$. 
$\circ$ and $\bullet$ indicate the sign and the amplitude of the wave function. 
The dashed line indicates how the honeycomb lattice is wound into a tube. 
(b) A typical current distribution which is 
resonant to the state illustrated in (a). 
The size of the colored arrows indicates the 
magnitude of the current density, while the black arrows indicate the source and 
drain electrodes.} 
\end{center}
\end{figure}

Equation (\ref{M2}) contains a factor 
${\rm Im}\{[\phi_\alpha^\ast(s)\phi_\alpha(d)]^2\}$, which gives $M_{\alpha} \propto 
\sin(2k_\perp^\alpha R\varphi)$ 
if we assume, for a heuristic purpose, a 
simple plane-wave form around the circumference 
$\phi_\alpha(s) = \phi_\alpha^\ast(d) = $
$|\phi_\alpha(s)| \exp(ik_\perp^\alpha 
R\varphi/2)$, 
where $k_\perp^\alpha$ is the wave number around the tube of the state 
$\alpha$, $R$ is the radius of the tube, and 
$\varphi$ is the angle subtended by the two electrodes, 
as shown in Fig. \ref{cylinder}(a).  
Only when the electrodes are attached asymmetrically ($\varphi \neq 0$) 
does the loop current arise, which 
immediately resolves a puzzle one might have on 
the symmetry: how a 
specific sense of rotation of the circulating current 
can arise when the 
cylindrical conductor has no structural chirality (as 
in zigzag or armchair CNTs).  

The above expression for $\Vec{M}$ contains another factor, 
${\rm Im}[\phi_\alpha(i)\phi_\alpha^\ast(j)]
\propto\sin\{k_\perp^\alpha[\theta(i)-\theta(j)]\}$, 
where we set $\phi_\alpha(\Vec{r})=|\phi_\alpha(\Vec{r})|
e^{ik_\perp^\alpha\theta(\Vec{r})}$ with $\theta$ the coordinate 
along the circumference. 
This factor expresses that loop currents are generated 
due to interference between the doubly degenerate states. 
Since these resonant states 
(encircled in Fig. \ref{band}) that participate 
in electronic transport lie near $\varepsilon_F=0$, 
the range of the value of $k_\perp$ is determined by the band structure 
of the conductor in general. In particular, the metallic CNTs with 
$(n_1-n_2)\equiv 0\;({\rm mod}\;3)$ have\cite{Reich} (see Fig. \ref{band}) 
\begin{eqnarray*}
k_\perp = \left\{
\begin{array}{ll}
0, & (n_1-n_2)/n\equiv 0\;({\rm mod}\;3):\; {\rm armchair\;CNT}\\
\pm \frac{2\pi}{3a}, & (n_1-n_2)/n\not\equiv 0\;({\rm mod}\;3):\; {\rm zigzag\;CNT},
\end{array}
\right.
\end{eqnarray*}
where $n$ is the 
greatest common divisor of $(n_1,n_2)$ and $a$ the length of the 
translational vector along the circumference. 
Thus, armchair CNTs with $k_\perp = 0$ should have 
nearly zero magnetic moments, which is contrasted with
zigzag CNTs for which magnetic moments are large. 
Chiral CNTs can have either $(n_1-n_2)/n \equiv 0$ or 
$\not\equiv 0\;({\rm mod}\;3)$, so even chiral structures, 
which may naively seem to guarantee a molecular solenoid, 
can have large moments only when the dispersion is correct.

So we conclude that large loop currents arise when we have 
(i) time-reversed pair of degenerate states, 
(ii) asymmetrically attached source and 
drain electrodes, and (iii) $k_{\perp}\neq 0$.  

\begin{figure}[htbp]
\begin{center}
\includegraphics[width=5cm]{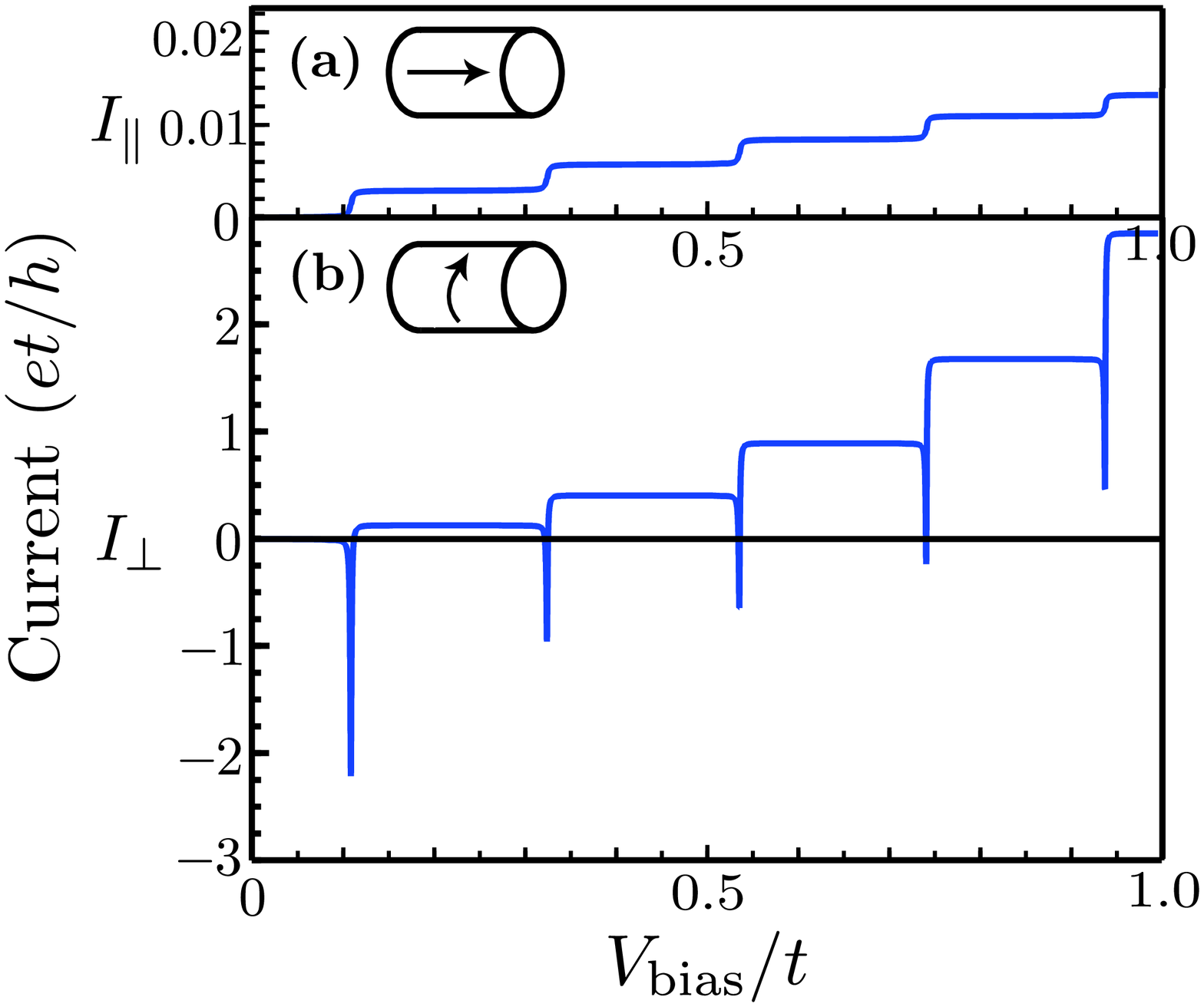}
\caption{\label{current3}(Color online) Calculated total current 
flowing (a) along the axis and (b) around the tube 
versus the bias voltage in a zigzag CNT, for which we have chosen 
smaller $(n_1,n_2)=(3,0)$ for clarity with $L=20a_0/\sqrt{3}$ 
with the same asymmetry ($R\varphi$) in the electrodes as in 
Fig. \ref{zigzag}(b). 
Note the different vertical scales.}
\end{center}
\end{figure}

\section*{NUMERICAL RESULTS FOR CARBON NANOTUBES} 
We now move on to the numerical calculation to confirm the 
analytical discussions above.  We have calculated the current and the 
magnetic moment for various types of CNT 
directly from Eq. (\ref{J1}) for 
$t'/t=0.2$ and $t''/t=2$ with $\lambda=0.02$.
Figure \ref{zigzag}(a) 
shows typical doubly degenerate eigenstates in a zigzag CNT, while 
the current density resonant to 
the state is depicted in Fig. \ref{zigzag}(b). 
We can immediately see that the current 
circulating around the tube is much larger than the one flowing along 
the axis, which endorses the discussion above.

Figure \ref{current3}(a) shows the total current flowing along the axis of 
the CNT as the function of the bias voltage. We can see that the current 
along the axis is nearly discretized, which is due to the resonant tunneling 
from a 1D electrode 
through the discrete electron levels existing at regular intervals in a CNT.  
Figure \ref{current3}(b) shows the total current circulating around the tube 
versus bias voltage, which is indeed larger than the current along the axis 
with a ratio $\sim \lambda^{-1} \gg 1$. 
The behavior consists of steps, but this time the increment of 
the current is not constant. 
This is attributed to the fact that the 
contribution to the current (or the orbital magnetic moment) from a 
pair of resonant eigenstates $\alpha$ differs 
from pair to pair [$\propto \varepsilon_\alpha$, Eq. (\ref{M2}), 
for 1D electrodes]. 

\section*{EVALUATION OF THE MAGNETIC MOMENT}
Let us finally estimate the order of magnitude of the orbital magnetic 
moment in CNTs.  From Eq. (\ref{M2}), 
$M_\alpha \sim (e/\hbar)\varepsilon_\alpha Ra_0$. 
Around the Fermi energy, a CNT has a 
cone-like dispersion with $\varepsilon_\alpha = 
[1\;{\rm meV}/L\;({\rm \mu m})]\times\alpha\;(\alpha=1,2,\ldots)$. 
Summing over the energy levels between $\mu_s=V_{\rm bias}$ 
and $\mu_d=0$, we obtain 
$M = \sum_\alpha M_\alpha \sim 10^{-3}\mu_B\times 
L\;({\rm \mu m})\times [V_{\rm bias}\;({\rm meV})]^2\times(R/a_0)$.
This amounts to $M \sim 10\mu_B $
for $L\sim 1\;;{\rm \mu m}$, $V_{\rm bias}\sim 30$ meV, 
and $R \sim 10a_0$.  
We also consider the magnetic field $B_{\rm ind}$ generated by the 
circulating currents.  We must satisfy 
$\Vec{M}\cdot \Vec{B}_{\rm ind}=M(\mu_0/2\pi R)(M/\pi R^2)\ll  \lambda t$ 
for the energy shift due to this to be negligible.  
If we combine this with the previous condition, 
we have to have
$[L\;({\rm \mu m})]^2(R/a_0)^{-1}$
$[V_{\rm bias}\;({\rm eV})]^4\ll \lambda\ll 10^{-3}
 [L\;({\rm \mu m})]^{-1}$ 
for $t\sim 1\;{\rm eV}$.
For $V_{\rm bias}$ and $R$ assumed above, the above set of 
inequalities is satisfied for $L\lesssim 1\;{\rm \mu m}$.

\section*{DISCUSSION} 
Here, we have concentrated on the one-body problem. The effects of 
the electron-electron interaction appear in, e.g., 
Coulomb blockade\cite{Bockrath,Tans} and Tomonaga-Luttinger liquid state.\cite{Bockrath2} 
So we need to take the interaction into account 
to describe the real systems more 
accurately, which is a future problem. 
Another point is that, although the self-induced magnetic field is small 
as mentioned above, some CNTs have large 
magnetic susceptibility,\cite{Ajiki-Ando} so its effect 
may be an interesting problem.
Also assumed is that each electrode touches only one atom of the CNTs. 
We have checked that the loop current is not significantly 
affected even when each electrode touches more than one atoms 
in the CNT.

Although orbital magnetic moments of CNTs come to be detected 
indirectly from the shift of energy levels in external magnetic fields, 
some direct experimental observation is desirable.  
We may be able to use the STM\cite{Wildoer&Odom} to probe the loop current 
or attach electrodes to CNT ropes or ``forest"\cite{hata} 
to detect the overall magnetization.  For the latter, some 
statistical average may have to be involved. 

\section*{ACKNOWLEDGMENTS}
We wish to thank illuminating discussions with Seigo Tarucha, 
Tsuneya Ando, and Takashi Oka.  This work was in part supported by 
a Grant in Aids for Creative Scientific Research Project 
from the Japanese Ministry of Education.


\begin{thebibliography}{30}
\item[]
*Present address: Institute for Solid State Physics, University of Tokyo, 
Kashiwa, Chiba 277-8582, Japan.
\bibitem{Saito}R. Saito, G. Dresselhaus, and M. S. Dresselhaus, {\it 
	Physical Properties of Carbon Nanotubes} (Imperial College Press, 
	London, 1998).
\bibitem{Reich}S. Reich, C. Thomsen, and J. Maultzsch, {\it Carbon 
	Nanotubes: Basic Concepts and Physical Properties} (Wiley-VCH, Berlin, 2003).
\bibitem{Hamada}N. Hamada, S. I. Sawada, and A. Oshiyama, Phys. Rev. Lett. {\bf 68}, 1579 (1992).
\bibitem{Saito2}R. Saito
, M. Fujita, G. Dresselhaus, and M. S. Dresselhaus,
Phys. Rev. B {\bf 46}, 1804 (1992).
\bibitem{Bockrath}M. Bockrath {\it et al.},
Science {\bf 275}, 1922 (1997).
\bibitem{Tans}S. J. Tans {\it et al.},
Nature (London) {\bf 386}, 474 (1997).
\bibitem{Minot}E. D. Minot {\it et al.},
Nature (London) {\bf 428}, 536 (2004).
\bibitem{Coskun}U. C. Coskun {\it et al.},
Science {\bf 304}, 1132 (2004).
\bibitem{Zaric}S. Zaric {\it et al.},
Science {\bf 304}, 1129 (2004).
\bibitem{Miyamoto}Y. Miyamoto, S. G. Louie, and M. L. Cohen,
Phys. Rev. Lett. {\bf 76}, 2121 (1996).
\bibitem{Yevtushenko}O. M. Yevtushenko
, G. Y. Slepyan, S. A. Maksimenko, A. Lakhtakia, and D. A. Romanov,
Phys. Rev. Lett. {\bf 79}, 1102 (1997);
G. Y. Slepyan
, S. A. Maksimenko, A. Lakhtakia, O. M. Yevtushenko, and A. V. Gusakov,
Phys. Rev. B {\bf 57}, 9485 (1998).
\bibitem{Ajiki-Ando}H. Ajiki and T. Ando, J. Phys. Soc. Jpn. {\bf 62}, 2470 (1993).
\bibitem{Ando}T. Ando, J. Phys. Soc. Jpn {\bf 74}, 777 (2005).
\bibitem{Lu}J. P. Lu, Phys. Rev. Lett. {\bf 74}, 1123 (1995).
\bibitem{Marganska}M. Marga\'{n}ska, M. Szopa and E. Zipper, Phys. Rev.	B {\bf 72}, 115406 (2005).
\bibitem{Datta}S. Datta, {\it Electronic Transport in Mesoscopic Systems} 
	(Cambridge University Press, Cambridge, 1995).
\bibitem{Nakanishi}S. Nakanishi and M. Tsukada, Jpn J. Appl. Phys., Part 
			 2 {\bf 
	37}, L1400 (1998); Surf. Sci. {\bf 438}, 305 (1999); Phys. Rev. 
       Lett. {\bf 87}, 126801 (2001).
\bibitem{Tsukada}M. Tsukada {\it et al.},
J. Phys. Soc. Jpn. {\bf 74}, 1079 (2005).
\bibitem{Hatano}K. Sasada and N. Hatano, Physica E (Amsterdam) {\bf 29}, 609 (2005).
\bibitem{Fujita}K. Nakada
, M. Fujita, G. Dresselhaus, and M. S. Dresselhaus,
Phys. Rev. B {\bf 54}, 17954 (1996).
\bibitem{Bockrath2}M. Bockrath {\it et al.},
Nature (London) {\bf 397}, 598 (1999).
\bibitem{Wildoer&Odom}J. W. G. Wild\"{o}er {\it et al.},
Nature (London) {\bf 391}, 59 (1998);
T. W. Odom {\it et al.},
{\it ibid.} {\bf 391}, 62 (1998).
\bibitem{hata}K. Hata {\it et al.},
Science {\bf 306}, 1362 (2004).
\end{thebibliography}

\end{document}